\documentclass[%
 reprint,
 amsmath,amssymb,
 aps,
pre,
]{revtex4-2}

\usepackage{graphicx}
\usepackage{dcolumn}
\usepackage{bm}
\usepackage{xcolor}

\begin{document}

\preprint{APS/123-QED}
\title{Polydispersity-driven dynamical differences between two- and three-dimensional supercooled liquids}

\author{Ilian Pihlajamaa}
\affiliation{%
Soft Matter \& Biological Physics, Department of Applied Physics,
Eindhoven University of Technology, P.O. Box 513, 5600MB Eindhoven, The Netherlands
}\author{Lotte S. van Gessel}
\affiliation{%
Soft Matter \& Biological Physics, Department of Applied Physics,
Eindhoven University of Technology, P.O. Box 513, 5600MB Eindhoven, The Netherlands
}\author{Corentin C.L. Laudicina}
\affiliation{%
Soft Matter \& Biological Physics, Department of Applied Physics,
Eindhoven University of Technology, P.O. Box 513, 5600MB Eindhoven, The Netherlands
}\author{Luc J. van Burik}
\affiliation{%
Soft Matter \& Biological Physics, Department of Applied Physics,
Eindhoven University of Technology, P.O. Box 513, 5600MB Eindhoven, The Netherlands
}\author{Liesbeth M.C. Janssen}
\affiliation{%
Soft Matter \& Biological Physics, Department of Applied Physics,
Eindhoven University of Technology, P.O. Box 513, 5600MB Eindhoven, The Netherlands
}

\date{\today}

\begin{abstract}

Previous studies have suggested a conundrum in the relaxation dynamics of polydisperse supercooled liquids. It has been shown that in two dimensions, the relative relaxation times of particles of different sizes become more similar as the material is cooled, whereas the opposite happens in three dimensions: they decouple.
Here we resolve this conundrum. First, we show that the coupling observed in two dimensions is an artifact of cage correction introduced to account for Mermin-Wagner fluctuations. Instead, the relative relaxation time of small and large particles in two dimensions remains constant or slightly decouples with temperature, as opposed to the substantial decoupling observed in three dimensions. Investigating these dimensional differences further,  we find through mobile cluster analysis that small particles initiate relaxation in both dimensions. As the clusters grow larger, they remain dominated by small particles in three dimensions whereas in two cluster growth becomes particle-size agnostic. We explain these findings with a minimal model by studying the distributions of single-particle barrier heights in the system, showing there is a clear difference in the environments of small and large particles, depending on the dimensionality. These findings highlight the critical role of dimensionality in glass formation, providing new insights into the mechanisms underlying the glass transition in polydisperse supercooled liquids.
\end{abstract}

\maketitle

\section{Introduction}

The liquid-to-glass transition is a ubiquitous phenomenon that can occur in virtually all liquids \cite{binder2011glassy, berthier2011theoretical}. As the system approaches the glass state, the material undergoes a drastic increase in viscosity (or structural relaxation time) with little obvious underlying structural changes. Empirically, the first stages of the viscosity growth are well described by a power law, which crosses over to an Arrhenius scaling under deeply supercooled conditions \cite{berthier2011theoretical, mallamace2010transport, mallamace2013dynamical}. 
Furthermore, the glass transition is accompanied by a host of other universal characteristics such as dynamic heterogeneity and Stokes-Einstein breakdown \cite{ediger1996supercooled, angell2000relaxation, berthier2011theoretical, cavagna2009supercooled, debenedetti2001supercooled, lubchenko2007theory}. However, at present the main microscopic mechanisms leading to these phenomena are still not well understood.

Many theories have been proposed to describe the dynamical crossover of the scaling of the relaxation time \cite{berthier2011theoretical}. The prevailing picture stemming from these theories is that so-called activated events emerge around the crossover point and start dominating the structural relaxation. However, different theories propose different physical interpretations of these events. Although there is a broad consensus that activated events involve the overcoming of barriers in the free-energy landscape, the spatio-temporal signature of these events remains disputed.
For instance, the random first-order transition scenario incorporates the idea of growing entropic droplets that form cooperatively rearranging regions of particles \cite{biroli2012random,  kirkpatrick2015colloquium, kirkpatrick1987dynamics, lubchenko2003barrier}. These regions grow in size with the degree of supercooling \cite{starr2013relationship, biroli2023rfot}. Alternatively, dynamic facilitation theory suggests a more kinetic picture in which sudden local structural rearrangements trigger others \cite{speck2019dynamic, chandler2010dynamics}. 
The practical difficulty in disentangling these different physical pictures lies in accessing the appropriate resolution and scale of real-space particle trajectories in the deeply supercooled regime,  both in experiments and in computer simulations.

Nevertheless, recent methodological breakthroughs in the equilibration of supercooled liquids have enabled the venture in the very low-temperature regime in silico \cite{grigera2001fast, brumer2004numerical, ninarello2017models, berthier2016equilibrium, berthier2019efficient, guiselin2022microscopic,scalliet2022thirty, berthier2017configurational, ozawa2017does, berthier2023modern}. This has provided important new insights into the deeply supercooled regime, including arguably the most detailed picture to date on the microscopic dynamics of a model glassformer \cite{scalliet2022thirty}. However, these state-of-the-art simulation methods are most efficient when the degree of size polydispersity is large, far exceeding the amount required to avoid crystallization. Although this is not problematic per se, it is important to understand the effects that the added polydispersity introduces to distinguish which observations are universal to all structural glassformers and which are model specific. 

In our recent work \cite{pihlajamaa2023influence}, we have demonstrated that the inherent size polydispersity of this model glassformer leads to a dynamic decoupling of small and large particles in three dimensions (3d). Interestingly, \citet{tong2023emerging} have demonstrated that the same mixture in two dimensions (2d) does not exhibit this decoupling. Instead, they have reported the opposite effect: The dynamics of large and small particles become more similar as the temperature is lowered. Spatial dimensionality thus appears to play an important role in the structural relaxation of polydisperse supercooled liquids, in particular regarding particle-size-resolved quantities. 

While it is well known that the physics of two- and three-dimensional systems can drastically differ in general \cite{strandburg1988two, bernard2011two}, it is commonly understood that this is not the case in amorphous solids \cite{harrowell2006glass, tarjus2017glass}. More specifically, long-wavelength fluctuations, known as Mermin-Wagner fluctuations, prevent the emergence of true long-range order at finite temperatures in two dimensions \cite{mermin1968crystalline}. In amorphous matter, these fluctuations significantly affect the single-particle trajectories as the temperature is lowered, without contributing to structural relaxation \cite{flenner2015fundamental, vivek2017long, illing2017mermin, shiba2016unveiling}. To quantify structural relaxation in a two-dimensional supercooled liquid, one must therefore correct for these effects by measuring the dynamical behavior of the particles relative to their direct surroundings. This can be done directly by computing cage-corrected quantities \cite{shiba2018isolating, tong2023emerging}, or indirectly by describing the correlations in terms of particle bonds instead of positions \cite{scalliet2022thirty}. Once appropriately corrected, supercooled liquids in two and three dimensions have been shown to share the same phenomenology \cite{harrowell2006glass, tarjus2017glass}. This is, however, seemingly at odds with the results of  \citet{tong2023emerging} and our findings \cite{pihlajamaa2023influence}. These observations raise fundamental questions on how both polydispersity and spatial dimensionality affect the structural relaxation of supercooled liquids at the microscopic level.

\begin{figure*}[ht]
    \centering
    \includegraphics[width = 0.9 \linewidth]{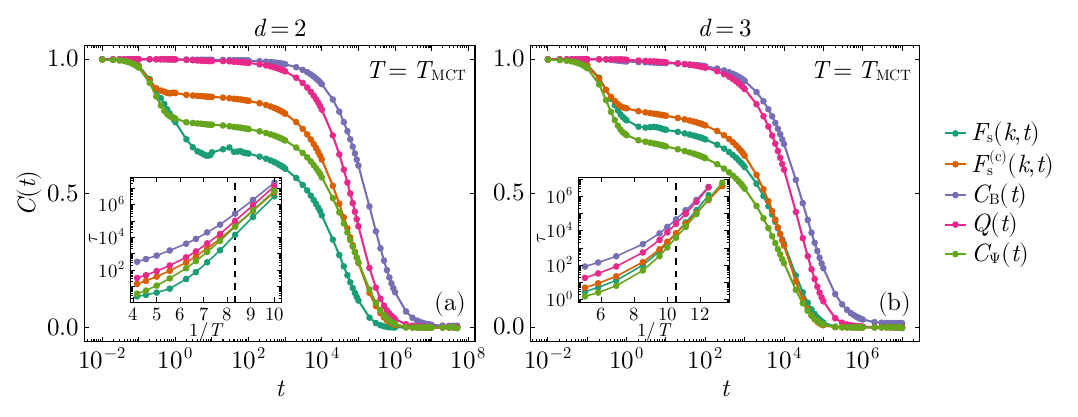}
    \caption{Different structural correlation functions $C(t)$ as a function of time at the mode-coupling crossover temperature in two (a) and three (b) dimensions. We show the self-intermediate scattering function $F_\mathrm{s}(k,t)$, the cage-corrected self-intermediate scattering function $F_\mathrm{s}^\mathrm{(c)}(k,t)$, the bond-breaking correlation function $C_\mathrm{B}(t)$ and the bond-orientational correlation function $C_\Psi(t)$. The latter is only shown in two dimensions. The scattering functions are computed at the peak of the structure factor ($k=6.8$ and $k=7.0$ in $d=2$ and $d=3$ respectively). The insets show the corresponding relaxation times, obtained by $C(\tau) = 1/e$ as a function of the inverse temperature. The vertical dashed line indicates the mode-coupling temperature $T_\mathrm{MCT}=0.12$ and $T_\mathrm{MCT}=0.095$ in $d=2$ and $d=3$ respectively. }
    \label{fig:1}
\end{figure*}

In this work, we rationalize the contradictory difference between two- and three-dimensional relaxation of polydisperse glassformers. We first show that the absence or presence of the dynamic decoupling is remarkably dependent on the cage-correction procedure, revealing that size-resolved effects can be masked by this protocol. We then provide a comprehensive analysis of size-resolved dynamics across dimensions, unifying earlier results from the literature \cite{tong2023emerging, pihlajamaa2023influence}. Our results unambiguously demonstrate qualitative differences between the two- and three-dimensional mechanisms governing structural relaxation in polydisperse supercooled liquids. Lastly, we explain the observed differences by showing that small particles have very different potential energy surroundings in two and three dimensions. 

\section{Results and Discussion}

\subsection{Average structural relaxation}

\begin{figure*}
    \centering
    \includegraphics[width=0.9\textwidth]{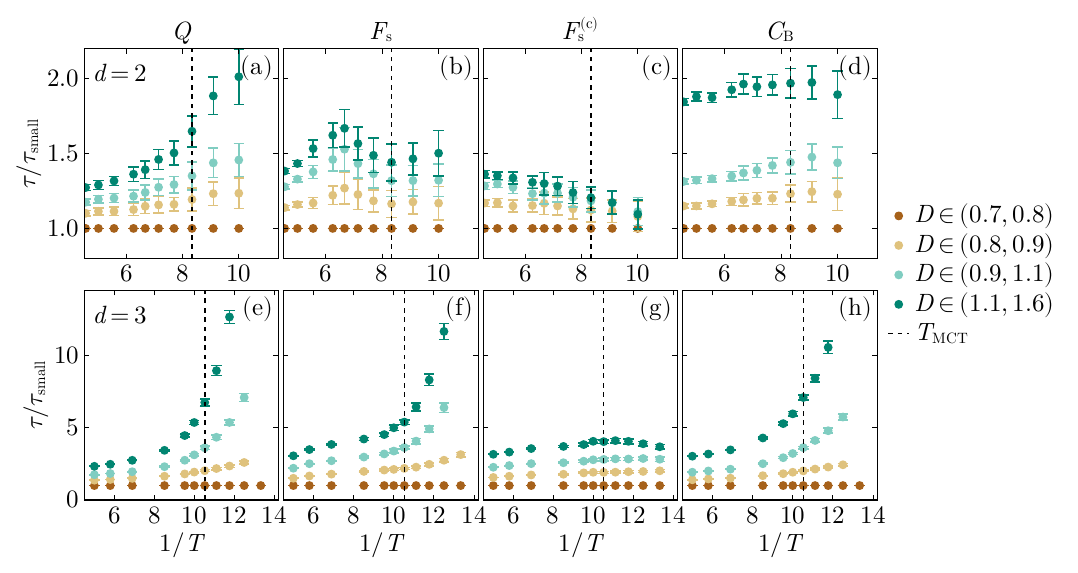}
    \caption{Relaxation times of the four different particle size quartiles relative to the smallest one as a function of the inverse temperature. The relaxation times are computed with the overlap function $Q$ in panels (a,e), with the self intermediate scattering function $F_s$ in panels (b,f), with the cage-corrected intermediate scattering function $F_\mathrm{s}^\mathrm{(c)}$ in panels (c,g), and with the bond-breaking correlation function $C_\mathrm{B}$ in panels (d,h). Panels (a--d) correspond to the two-dimensional system and (e--h) to the three-dimensional one. The black dashed line indicates the mode-coupling temperature.}
    \label{fig:2}
\end{figure*}

Before comparing particle-size resolved relaxation across dimensions $d$, it is helpful to first quantify the overall structural relaxation in the respective systems. 
For this, we may consider various time-dependent correlation functions, say $C(t)$, to quantify the relaxation dynamics. These include the self-intermediate scattering function $F_\mathrm{s}(k,t) = \langle \sum_i \cos \left(\textbf{k}\cdot(\textbf{r}_i(t)-\textbf{r}_i(0)\right)\rangle/N$ at wave number $k = |\textbf{k}|$ and the overlap function $Q(t)= \langle \sum_i \Theta (a - |\textbf{r}_i(t)-\textbf{r}_i(0)|)\rangle/N$, which both probe correlations between particle positions $\textbf{r}_i$ at different times $t$. Here, $\Theta(a-x)$ is some window function that goes from 1 to 0 around $x=a$. Alternatively, one can quantify structural correlation using information about nearest-neighbor bonds. Examples are the bond-orientational correlation function and the bond-breaking correlation function. The former is defined as $C_{\Psi}(t) = \langle\sum_i \Psi_i(t)\Psi_i^*(0)\rangle/\langle\sum_i\Psi_i(0)\Psi_i^*(0)\rangle$, which measures the evolution of local symmetry of the neighborhood characterized by a local bond-orientational order parameter $\Psi_i$. The bond-breaking correlation function is defined as $C_\mathrm{B}(t) = \langle\sum_i n_i(t|0)/n_i(0)\rangle/N$ and measures the loss of nearest neighbors in time. Here, $n_i(t_2|t_1)$ is the number of particles that are neighbors of particle $i$ at both $t=t_2$ and $t=t_1$. Each of these correlation functions is a well-established quantity and has been shown to describe the time scale over which equilibrium density fluctuations decay \cite{flenner2015fundamental}. 

In two-dimensional systems, density fluctuations include not only structural changes, which we aim to quantify, but also long-wavelength Mermin-Wagner (MW) fluctuations that do not contribute to structural relaxation and the viscosity. While the correlation functions based on neighbor bonds ($C_\mathrm{B}$ and $C_\Psi$) are by construction blind to these MW fluctuations, those that depend explicitly on particle positions are influenced by them.  To remedy this, one may consider `cage-corrected' functions where the motion of the center of mass of the set of neighbors is subtracted \cite{shiba2018isolating, tong2023emerging}. For example, the cage-corrected self-intermediate scattering function may be defined as 
\begin{equation}
    F_\mathrm{s}^{\mathrm{(c)}}(\textbf{k},t)= \frac{1}{N}\left< \sum_i \cos \left[\textbf{k}\cdot\left(\Delta \textbf{r}_{i}(t)-\Delta\textbf{R}_i^{\mathrm{CM}}(t)\right)\right]\right>,
\end{equation}
where $\Delta \textbf{r}_{i}(t) = \textbf{r}_i(t) - \textbf{r}_i(0)$, $\Delta\textbf{R}_i^{\mathrm{CM}}(t) = \textbf{R}_i^{\mathrm{CM}}(t|0)-\textbf{R}_i^{\mathrm{CM}}(0|0)$ with $\textbf{R}_i^{\mathrm{CM}}(t|0) = \frac{1}{n_i(0)}\sum_{l=1}^{n_i(0)} \textbf{r}_l(t)$ denoting the position at time $t$ of the center of mass of the cage of particle $i$, with the cage defined at $t=0$.

In this work we quantify the structural relaxation in a recently developed continuously size-polydisperse mixture introduced by Berthier and coworkers \cite{ninarello2017models, scalliet2022thirty}. We simulate both two- and three-dimensional systems where the diameters of the particles are drawn from the same distribution with a polydispersity of 23\% (see Methods for details). 
For these systems, the respective mode-coupling temperatures are given by $T_\mathrm{MCT} = 0.12$ (2d) and $T_\mathrm{MCT} = 0.095$ (3d), where $T_\mathrm{MCT}$ marks the power-law-to-Arrhenius crossover temperature. Unless otherwise specified, we focus on structural relaxation on length scales corresponding to the first neighbor shell, i.e.,~ $k=6.7$ and $a=0.5$ in units of average particle diameter.

Figure~\ref{fig:1} compares the structural relaxation as quantified by the different correlation functions at $T_\mathrm{MCT}$, both in two and three dimensions. We observe that all correlation functions capture the slow dynamics of the system. Qualitatively, we observe that the intermediate scattering function and bond-orientational function exhibit a two-step decay, a known characteristic of caging behavior in supercooled liquids. In contrast, the bond-breaking and overlap correlation functions do not show an intermediate plateau, because they are not sensitive to intra-cage motion. Nevertheless, all are capable of capturing the dynamic crossover at $T_{\mathrm{MCT}}$ across dimensions, as shown in the corresponding insets. 

At the quantitative level, however, there are notable differences among the correlation functions. Specifically, the relaxation times associated with $F_\mathrm{s}(k,t)$ and $C_\mathrm{B}(t)$ differ by more than one order of magnitude, while the relaxation time of $Q(t)$ lies in between (see insets of Fig.~\ref{fig:1}). These trends can be explained by recognizing that the bond-breaking correlation function $C_\mathrm{B}(t)$ requires more structural reorganization to decay compared to $F_\mathrm{s}(k,t)$ and $Q(t)$, because the former is sensitive only to loss of neighbors and not merely particle displacements. A similar observation has been made by \citet{flenner2016dynamic} in a two-dimensional binary glassformer. 

Our results also show the validity of the cage-correction procedure to account for Mermin-Wagner fluctuations.  While the procedure affects the plateau height across dimensions,  in three dimensions the absence of MW fluctuations is reflected in the identical long-time decay of $F_\mathrm{s}(k,t)$ and $F^{\mathrm{(c)}}_\mathrm{s}(k,t)$. In contrast, the presence of MW fluctuations in two dimensions leads to a significant difference in the measured decorrelation times.
We additionally observe that the bond-orientational correlation function $C_\Psi(t)$ shares the relaxation time with the cage-corrected intermediate scattering functions $F^{\mathrm{(c)}}_\mathrm{s}(k,t)$ at low temperatures. From these results, cage correction of the intermediate scattering function therefore seems an appropriate way to remove the effects of MW fluctuations, while still quantifying the structural relaxation dynamics. 

\subsection{Size-resolved structural relaxation}
As mentioned in the introduction, seemingly conflicting results have been reported between our earlier work (Ref.~\cite{pihlajamaa2023influence}) and that of \citet{tong2023emerging} regarding the size-specific relaxation times in two and three dimensions. For the two-dimensional case, Ref.~\cite{tong2023emerging} shows that the relaxation times of the small and large particles, as measured by a size-resolved $F_\mathrm{s}^{\mathrm{(c)}}(t)$, become progressively similar as the temperature is decreased. In contrast, Ref.~\cite{pihlajamaa2023influence} reports that the dynamics of small and large particles progressively decouple in three dimensions (as measured by $F_\mathrm{s}(t)$ or $C_\mathrm{B}(t)$). In this section, we shed light on this discrepancy. Specifically, we report the size-resolved relaxation times as obtained from four different correlation functions, $Q(t)$, $F_\mathrm{s}(t)$, $F_\mathrm{s}^{\mathrm{(c)}}(t)$, and $C_\mathrm{B}(t)$. To compute particle-size dependent quantities, we sort the particles into four groups that correspond to the quartiles of the size distribution (see the Methods section for details). The remaining bond-orientational correlation function, $C_\Psi(t)$, is ill-suited for size-resolved quantities because its plateau height drastically depends on particle size, as intermediately sized particles have a local environment that is much more six-fold symmetric than that of large or small particles \cite{tong2023emerging}. 

The results of our size-resolved analysis are shown in Fig.~\ref{fig:2}, where we visualize the relaxation times relative to those of the particles in the smallest quartile as a function of the inverse temperature. Overall, looking at the relaxation times according to $Q$, $F_\mathrm{s}$, or $C_\mathrm{B}$, indeed, we find that the dynamical decoupling between small and large particles is much more evident in three than in two dimensions. 
More precisely, in three dimensions, the data from the overlap function, self-intermediate scattering function, and the bond-breaking correlation function show that, around the mode-coupling temperature, the relaxation times of the largest particles progressively increase relative to the smallest ones. At the lowest temperature studied, the relaxation times of the smallest and the largest quartile are separated by more than an order of magnitude. In two dimensions, this increasing separation is weakly present according to $Q(t)$, while the other functions show no clear increase in the relative relaxation times upon supercooling. Notably, no clear change is found around the mode-coupling point, and the relative relaxation time $\tau_\mathrm{large}/\tau_\mathrm{small}$ remains smaller than 2 within our temperature range [Fig.~\ref{fig:2}(a,b,d)]. 

However, the work of Ref.~\cite{tong2023emerging} reported that in two dimensions the size-resolved relaxation times converge towards each other upon supercooling. We find that this observation can only be made if the cage-corrected intermediate scattering function $F_\mathrm{s}^{\mathrm{(c)}}(k,t)$ is used to quantify the dynamics [as shown in Fig.~\ref{fig:2}(b)]. In the same vein, we find that the cage-correction procedure qualitatively affects the results in three dimensions. 
In particular, the relaxation time ratios become approximately constant after cage correction  [Fig.~\ref{fig:2}(e)], thus
masking the decoupling determined by the other correlation functions. This suggests that the cage correction procedure leads to artifacts in the size-resolved dynamics. Physically, we may also understand this as follows. 
The cage correction inherently incorporates the effects of neighboring particle movements, rendering the resulting function no longer a strictly single-particle property. Hence, the size-resolved $F_\mathrm{s}^{\mathrm{(c)}}(k,t)$ also implicitly probes the dynamics of other (neighboring) particle sizes. It is unclear why this effect is less pronounced in $C_\mathrm{B}(t)$, which by definition also incorporates the dynamics of neighboring particles.


As mentioned above, the bond-orientational correlation function $C_\Psi(t)$ should also be avoided for single-particle properties in this model system, as it shows significantly varying plateau heights depending on particle size.
This is caused by a size-dependent local cage symmetry due to the non-additivity, energetically favoring small and large particles to neighbor each other. This results in the fact that the particles of intermediate sizes have higher plateaus according to $C_\Psi(t)$ and thus longer relaxation times than both small and large particles, both in $d=2$ and $d=3$. We have verified that the qualitative aspects of our results hold if the polydispersity index is increased up to 39\% in two dimensions. This corresponds to a mixture with a maximum size ratio of 4, in which case the relative relaxation time $\tau_\mathrm{large}/\tau_\mathrm{small}\approx 3$ for the lowest temperatures investigated. 

In conclusion, the dynamical decoupling between small and large particles is very weak or absent in two dimensions, while it is strong in three. Furthermore, one should be careful which observable is used to quantify size-resolved dynamics, in particular related to Mermin-Wagner fluctuations. 
In the following, we elucidate the underlying mechanisms that lead to different structural relaxation in two- and three-dimensional polydisperse glassformers. 


\subsection{Polydispersity-driven dynamic heterogeneity}

\begin{figure}
    \centering
    \includegraphics[width=\linewidth]{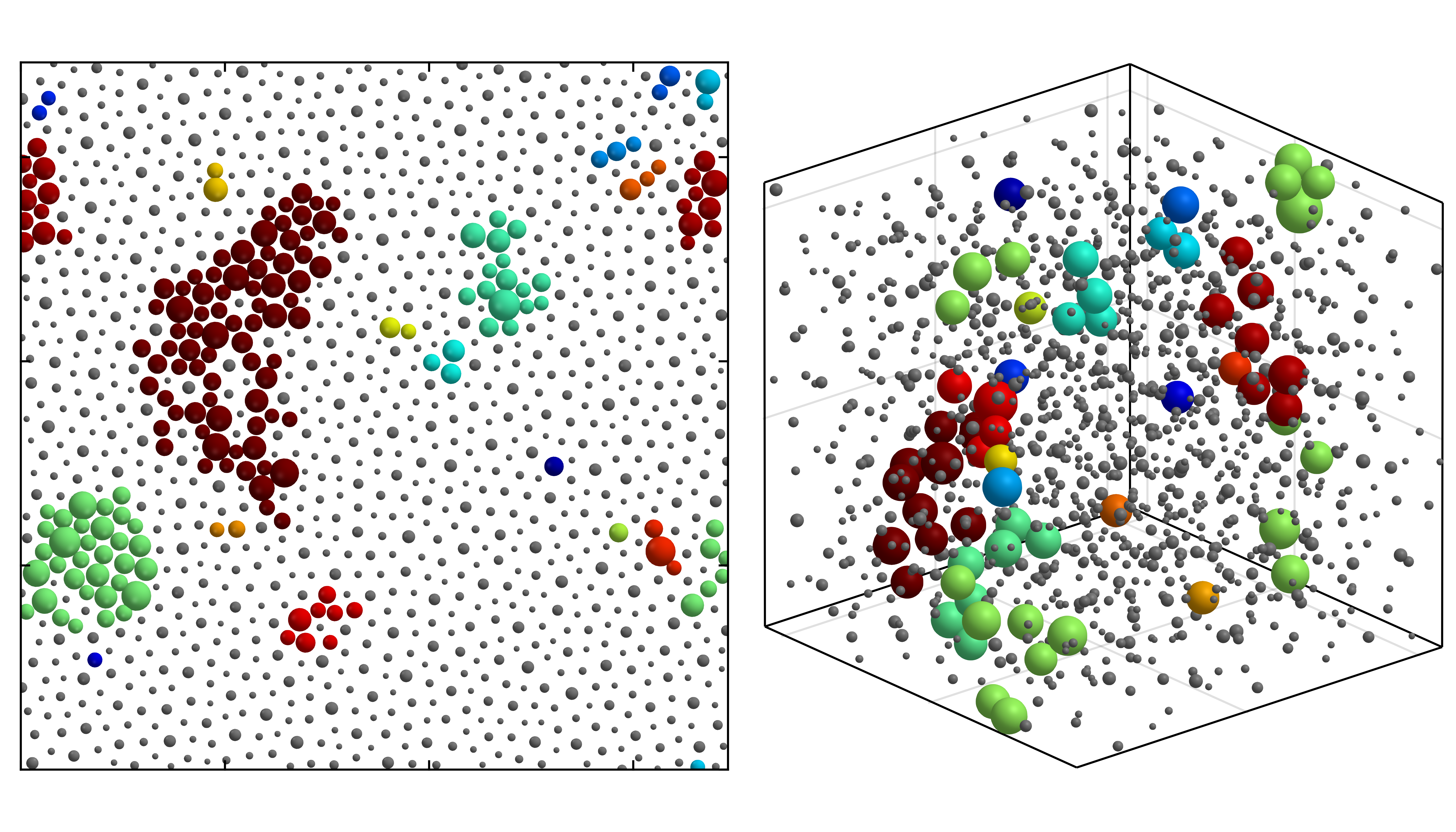}
    \caption{Clusters of mobile particles in two and three dimensions below the mode-coupling temperature. Each cluster is given a distinct color. Immobile particles are drawn small in gray. The left panel was created at temperature $T=0.11$ for $t=10^5$, and the right at $T=0.075$ and $t=8\times10^5$.}
    \label{fig:cluster}
\end{figure}

To gain deeper insight into the different relaxation dynamics of small and large particles across dimensions, we conduct a quantitative analysis of mobile clusters \cite{kob1997dynamical, donati1998stringlike, donati1999spatial}. To differentiate between mobile and immobile particles, we calculate the bond-breaking order parameter $C_\mathrm{B}^i(t)$ for each particle. This parameter represents the fraction of a particle's neighbors that are unchanged between time $t$ and time $0$. We define particle $i$  as \textit{mobile} if $C^i_\mathrm{B}(t) < 0.5$, meaning it has lost more than half of its neighbors and as \textit{immobile} otherwise. By spatially clustering the mobile particles, we create a detailed map highlighting the regions of the material that have undergone significant structural relaxation. Representative clusters in both two and three dimensions are illustrated in Fig.~\ref{fig:cluster}, providing a view of the relaxation behavior in different spatial dimensions.

\begin{figure}
    \centering
    \includegraphics[width=\linewidth]{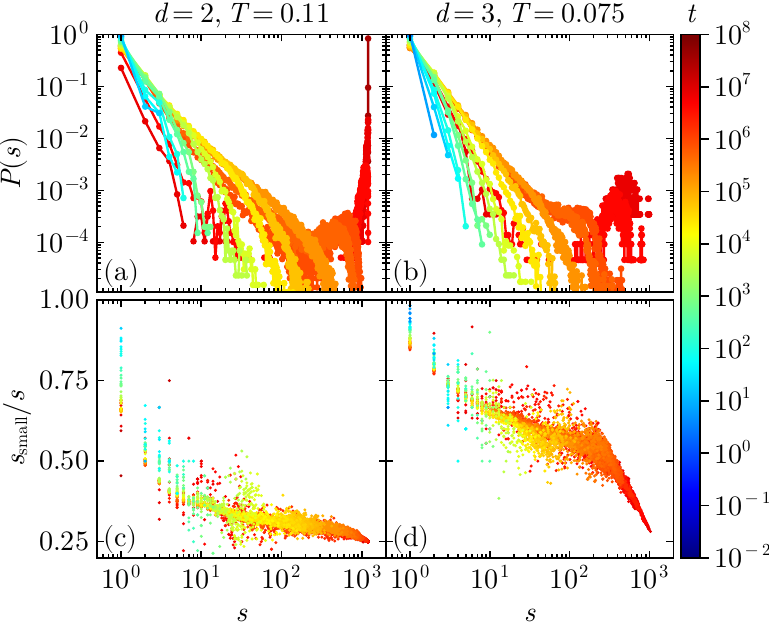}
    \caption{Cluster size distributions $ P(s) $ of mobile particles. Panel (a) shows $P(s)$ over different time intervals $t$ for dimension $ d = 2 $ at temperature $ T = 0.11 $. Panel (b) shows the same in three dimensions at $T=0.075$. Panels (c) and (d) show the fraction of particles in clusters of size $ s $ that are classified in the smallest quartile of the particle size distribution for the same respective conditions. The relaxation times are $\tau_\alpha = 1.8 \times10^6$ and $\tau_\alpha = 1.4 \times10^7$ in the chosen state points in two and three dimensions respectively. The colors denote the age of the clusters. The data reveal size-dependent behaviors of mobile particle clusters, with differences between the two-dimensional and three-dimensional systems at various cluster sizes $ s $. 
}
    \label{fig:clusterdist}
\end{figure}

To quantitatively investigate the relaxation of the material, we analyze the size distribution $P(s)$ of mobile clusters, where $s$ represents the number of particles within a given cluster. The evolution of $P(s)$ is depicted over time in Fig.~\ref{fig:clusterdist}(a,b) in two and three dimensions at temperatures below the mode-coupling temperature. At time scales shorter than the structural relaxation time, we observe that this cluster size distribution takes a power-law form $P(s)\propto s^{-\alpha}$, where the exponent $\alpha$ depends on the age $t$ of the clusters and the temperature $T$. The appearance of power-law behavior stems from the underlying percolation problem \cite{stauffer2018introduction}. As $t$ increases, the average cluster size grows until the clusters coalesce, and a percolating cluster emerges. This percolating cluster grows until it contains all $N=1200$ particles in the simulation. Overall, our findings show that the qualitative features of $P(s)$ are independent of dimensions and temperature (see Supplemental Information (SI) for different temperatures). This picture also aligns with previous reports in the literature, including a different cluster analysis on the same model \cite{scalliet2022thirty}, and on different ones \cite{mel1995long, gebremichael2001spatially}. 

It is natural to ask how particle size influences this cluster growth. To this end, we examine the fraction of small particles in clusters of size $s$. Specifically, we calculate the average proportion $s_\mathrm{small}/s$  of particles in each cluster that fall within the smallest size quartile. We show this fraction in Fig.~\ref{fig:clusterdist}(c,d) in two and three dimensions, respectively. In a system where size does not affect relaxation, this fraction would remain constant at 0.25, regardless of $d$, $s$, $t$, or $T$. However, our results reveal a strong dependence on particle size. Let us first focus on the smallest possible clusters, i.e., isolated mobile particles with $s=1$. Our results show that overall these individual mobile particles are highly likely to be small, with likelihoods of around $75\%$ in $d=2$ and $90\%$ in $d=3$. 
Moreover, the earliest single-particle relaxation events (indicated by blue scatter points in Fig.~\ref{fig:clusterdist}) are even more certain to involve a small particle, with $s_\mathrm{small}/s$ approaching 100\% for very short times.
There are two possible interpretations for this last observation. Either the majority of initial ``activated" events involve small particles, or initial activated events of larger particles must inherently involve more than one particle. From panels \ref{fig:clusterdist}(a) and (b), however, it is clear that the latter case is very rare, since at early times (blue data points) $P(s>1)$ is very small compared to $P(s=1)$. Thus, irrespective of dimension, our results show that very early structural relaxation is dominated by single, small particles. 

We focus now on larger mobile clusters ($s>1$), which naturally appear as relaxation progresses. For such larger clusters, we do find a stark difference between the two- and three-dimensional systems. In two dimensions, the small particle fraction drops quickly as cluster size increases [Fig.~\ref{fig:clusterdist}(c)], with larger clusters showing only a very slight excess of $25\%$. This proportional representation of small particles in the two-dimensional clusters suggests that particle size does not severely impact the propagation of structural relaxation. In contrast, in three dimensions [Fig.~\ref{fig:clusterdist}(d)], small particles continue to be overrepresented even in large clusters up to $s\approx100$. As the clusters grow further, the small particle fraction decreases, implying that the larger particles participate more. Proportional representation of $25\%$ is only achieved on time scales when essentially all particles are mobile ($s\approx 1000$). We have found that this dimensional difference appears already at higher temperatures, well above $T_\mathrm{MCT}$ (see SI). We have additionally verified that these observations are unchanged if a displacement criterion is used instead of $C_\mathrm{B}^i$ to classify the mobility of the particles.

\begin{figure}
    \centering
    \includegraphics[width=\linewidth]{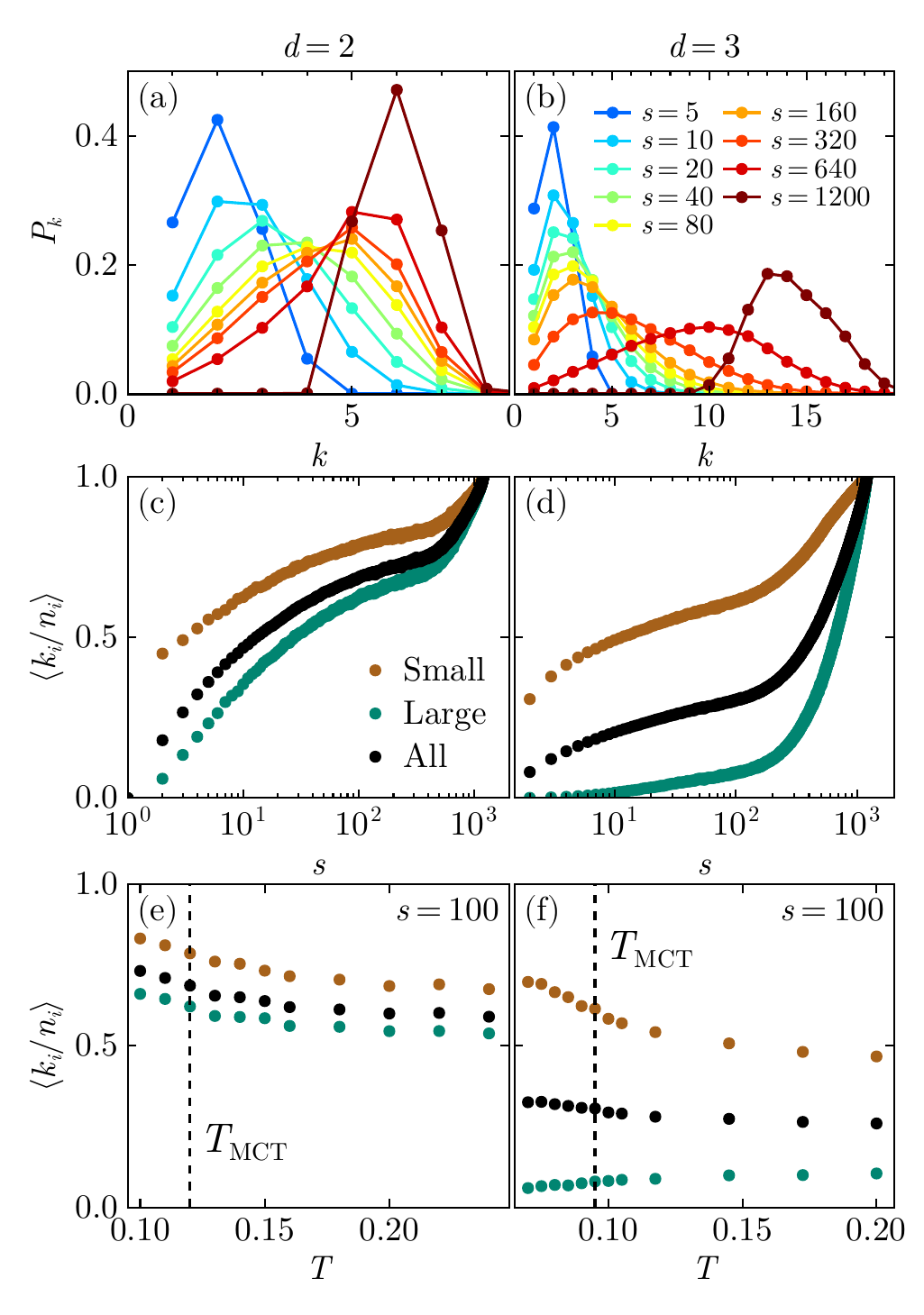}
    \caption{Connectivity within mobile clusters in two (a, c, e) and three (b, d, f) dimensions. Panels (a) and (b) show the degree distribution $P_k$ of the mobile clusters, that is, the probability that a particle in a cluster has $k$ neighbors that are also mobile (and thus also part of the cluster) for different cluster sizes $s$. Panels (c) and (d) show for a mobile particle in a cluster of size $s$ the average proportion of neighbors that are mobile (black circles). With blue and brown circles, the average proportions of respectively large and small mobile neighbors are indicated. For (a-d), the temperature is set to $T_\mathrm{MCT}$. In panels (e) and (f), we show the same as in (c, d) for a fixed cluster size $s=100$ while varying the temperature. The mode-coupling temperature is indicated with a vertical dashed line.
}
    \label{fig:connectivity}
\end{figure}

To understand the effect that this difference has on the morphology of the mobile clusters, we investigate their connectivity as a function of their size $s$.  In Fig.~\ref{fig:connectivity}(a,b), we show in two and three dimensions the degree distribution $P_k$, where the degree $k$ is the number of mobile neighbors of a mobile particle. These distributions provide insight into the structure of a mobile cluster: If the average degree is low, the clusters are string-like, whereas a large average degree implies the clusters are dense. The data of Fig.~\ref{fig:connectivity}(a,b) reveal that, both in two and three dimensions, small clusters are string-like, in correspondence with literature \cite{donati1998stringlike, gebremichael2004particle, chacko2024string} However, as the clusters grow beyond $s\sim20$ (and thus comprise multiple `fundamental' string-like excitations), the two-dimensional clusters compact, as evidenced by the increased average degree. In contrast, in the three-dimensional case, much larger clusters still remain string-like, and the peak of the distribution shifts to the right only when the clusters have grown to a size $s$ of over a hundred particles.

Although degree distributions illustrate the dynamics of cluster growth, they are ill-suited for a clear comparison across dimensions. This is because the number of neighbors $n_i$ for a typical particle $i$ varies significantly between two and three dimensions, with $\langle n_i\rangle = 6.0$ and $\langle n_i\rangle = 14.3$, respectively. Arguably a better observable is the average fraction $\langle k_i/n_i\rangle$ of mobile neighbors of a mobile particle $i$. This metric is presented in Fig.~\ref{fig:connectivity}(c) and (d) in two and three dimensions with black markers. The data confirm that there is a stark contrast between the two- and three-dimensional system: For clusters of size $s=100$, the clusters are significantly denser in $d=2$ (fraction of 0.7) than in $d=3$ (fraction of 0.3). 
Let us specialize these fractions on neighbors within a specific size quartile, i.e., we now focus on the ratio of the number of small mobile neighbors to small neighbors, regardless of $i$'s size. These data, indicated with the colored curves in Fig.~\ref{fig:connectivity}(c) and (d) show in three-dimensional space that small particles predominantly make up clusters in the intermediate stage, while large particles become involved only towards the final phases of structural relaxation. In contrast, for $d=2$ large and small particles do not exhibit this difference. 
Inspecting this dimensional difference as a function of temperature for $s=100$ in Fig.~\ref{fig:connectivity}(e) and (f), we observe that they are exacerbated by the dynamic decoupling in three dimensions and the absence thereof in $d=2$, even though the difference is already present at high temperatures.

Overall, our results paint the following picture: In both two and three dimensions, initial relaxation events in the material are mostly triggered by small particles undergoing highly localized structural reorganizations or cage escapes. As shown by \citet{scalliet2022thirty}, these initial events facilitate further relaxation events in their vicinity, resulting in string-like clusters both in two and three dimensions. Our work reveals that there is a clear dimensional difference in the mechanism by which this facilitation propagates as the clusters grow. In two dimensions, structural relaxation is more or less indiscriminate with respect to particle size, producing dense clusters. In three dimensions, however, subsequent relaxation events still highly favor small particles over larger ones, resulting in clusters that are very sparse. These differences highlight the nontrivial interplay of spatial dimension and polydispersity in the structural relaxation of dense fluids. 

\subsection{Size-resolved local energy landscapes}\label{sec:D}

The observation that relaxation events of single small particles predominantly initiate the growth of mobile clusters (Fig.~\ref{fig:clusterdist}(a-b)) motivates an in-depth analysis of single-particle processes. To understand the different observed behaviors of small and large particles across dimensions, we consider the energy landscapes as experienced by a tagged particle in the supercooled liquid. Because the interaction potential in this model is short-ranged, a given particle interacts only with its nearest neighbors who exert a caging effect on it. In this section, we investigate the effective energy barriers created by these cages while varying the size $D$ of a tagged particle, the temperature $T$, and the dimensionality $d$. 

\begin{figure}[ht]
    \centering
    \includegraphics[width=\linewidth]{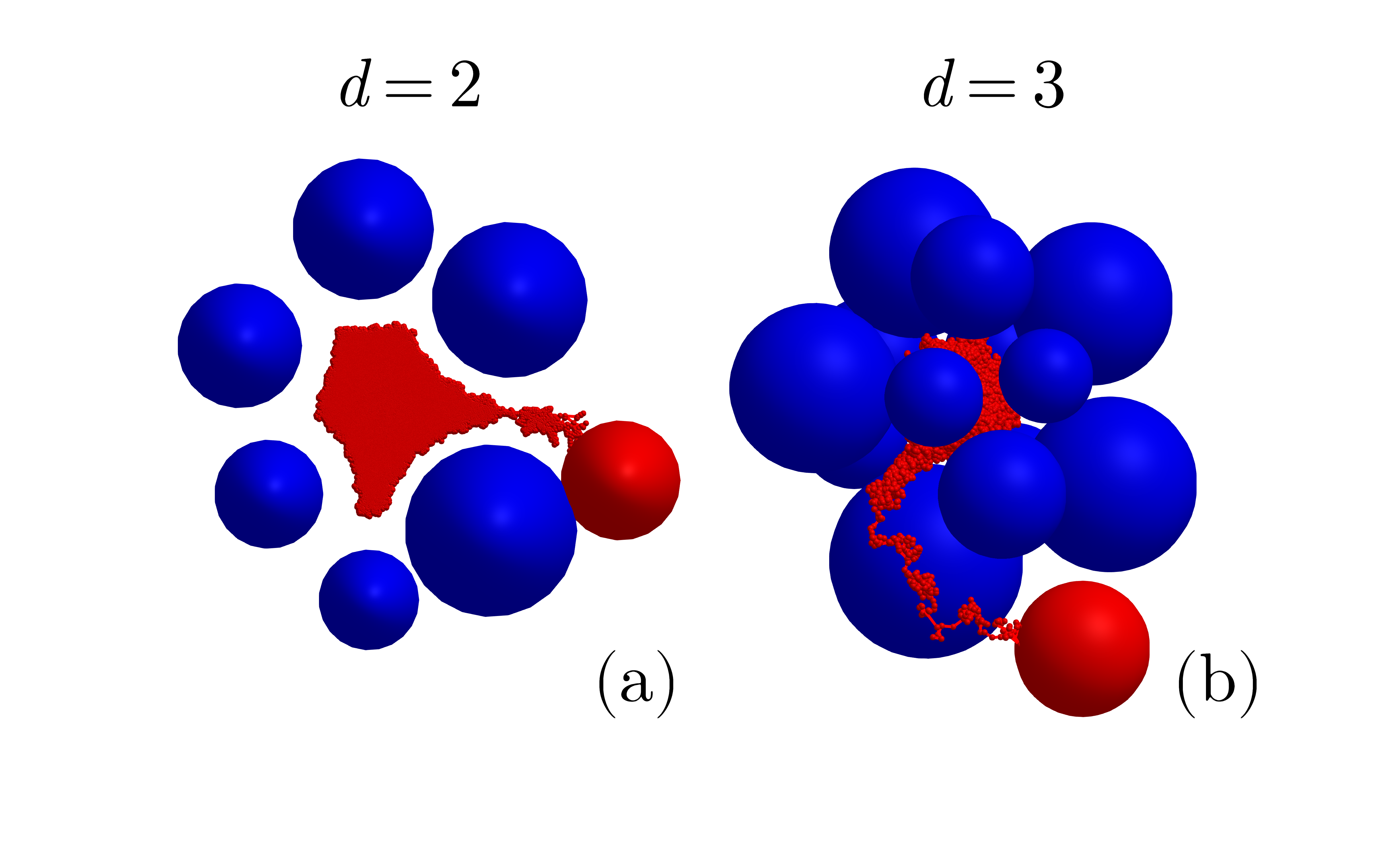}
    \caption{The Kramers escape problem: Panels (a) and (b) show the trajectory of a particle in its cage of pinned particles until it escapes in two and three dimensions respectively. The red particle is mobile and the blue ones are pinned, creating a static external field.}
    \label{fig:3}
\end{figure}

Briefly, our procedure is as follows (see Sec.~\ref{sec:methods} for details): We select a tagged particle and determine its neighbors from an equilibrium configuration of the model supercooled liquid. Coupling this particle to a heat bath, we consider its dynamics in the static external field as created by the neighboring particles from the equilibrium configuration. This framework reduces the system to a two- or three-dimensional version of the classical Kramers escape problem \cite{kramersproblem}. Figure \ref{fig:3} illustrates this model in both two and three dimensions. By sampling the positions of the central particle and its neighbors directly from equilibrated configurations (and not from e.g.~an inherent structure quench), we capture the correct cage distribution as experienced by the tagged particle in a supercooled liquid. The dynamic cage fluctuations, including those pertaining to cooperative rearrangements, enter the Kramers model as quenched disorder. 

For a tagged particle to escape its cage, it must overcome (at least) the minimal energy barrier $E_\mathrm{b}$ in the external field created by its neighbors. According to Kramers theory \cite{kramersproblem, van1992stochastic}, if $E_\mathrm{b}\gg k_\mathrm{B}T$, the average escape time $\tau_\mathrm{K}$ is described by an Arrhenius law  
\begin{equation}
    \tau_\mathrm{K} \propto e^{E_\mathrm{b}/k_\mathrm{B}T}.
    \label{eq:1Dkramer}
\end{equation}
The prefactor depends on the local curvature at the center of the cage and that at the potential barrier crossing point \cite{kramersproblem} Thus, the energy barrier height predominantly determines the dynamics in this model. This essentially recasts the question of the origin of different relaxation behaviors of small and large particles in terms of their respective energy barriers. 

We determine the energy barrier using two methods, which we refer to as the variable-angle method and the constant-angle method (see Sec.~\ref{sec:methods} for details). They respectively provide lower and upper bounds for the true minimal energy barrier. These methods are considerably simpler than perhaps more commonly used ones, such as the nudged elastic band method \cite{sheppard2012generalized} and other saddle point localization schemes \cite{muller1979location}, which are also applicable in higher-dimensional transition state problems. We have verified that our results are in agreement with those obtained from the nudged elastic band procedure.

\begin{figure}[ht]
    \centering
    \includegraphics[width = \linewidth]{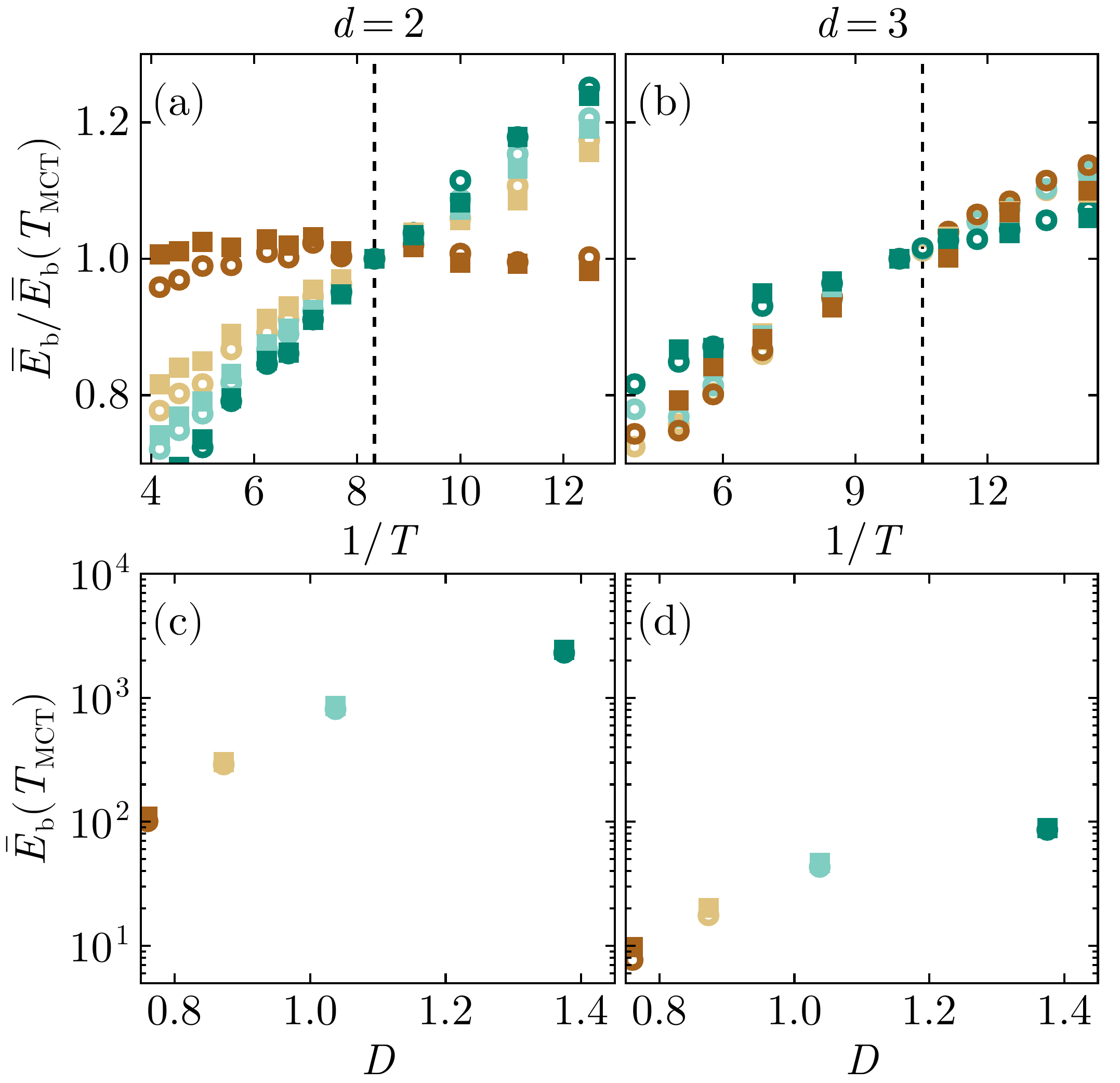}
    \caption{Average size-resolved single-particle energy barriers in two and three dimensions. Panels (a) and (b) show the diameter-resolved barrier heights as a function of the inverse temperature in two and three dimensions. The barrier heights are normalized by the barrier height at $T_\mathrm{MCT}$ to collapse the different diameter bins. Panels (c) and (d) show the barrier heights at $T_\mathrm{MCT}$ per particle diameter quartile. Legend colors are the same as in Fig.~\ref{fig:2}. The open circles correspond to barrier heights determined with the variable angle method whereas the squares correspond to the constant angle method.
    }
    \label{fig:4}
\end{figure}

In Fig. \ref{fig:4}(a,c), we present the average barrier height $\overline{E}_\mathrm{b}$ as a function of the inverse temperature in two and three dimensions for particles of different sizes. Here, the average is taken over different particles and equilibrium configurations, i.e.~it is the disorder average in the Kramers model. In order to admit a fair comparison across dimensions, we have normalized the barrier heights by their respective values at the MCT temperature, $\overline{E}_\mathrm{b}(T_\mathrm{MCT})$, the latter of which are shown in Fig.~\ref{fig:4}(b,d). Results from both methods are included, showing that they are consistent with each other.

Our results in Fig. \ref{fig:4}(a,c) reveal that, overall, as the temperature is reduced from the onset of the supercooled regime down to and across $T_\mathrm{MCT}$, the average barrier height increases in both two and three dimensions. The relative increase of the average energy barrier $\overline{E}_\mathrm{b}/\overline{E}_\mathrm{b}(T_\mathrm{MCT})$ remains roughly of the same magnitude across dimensions. This growth of the energy barriers must ultimately follow from subtle changes in the average cage structure, consistent with established structural changes found upon supercooling \cite{royall2015role, zhang2020revealing, pihlajamaa2023emergent}. However, when considering the size-resolved energy barriers, an exception is found for the smallest particles in two dimensions. For these, we instead find that the average barrier height remains virtually temperature-independent, even well below the mode-coupling temperature. 

The \textit{absolute} average energy barriers $\overline{E}_\mathrm{b}$, however, exhibit a striking difference between dimensions. Specifically, as shown in Fig.~\ref{fig:4}(b,d), the barrier heights in $d=3$ are at least one order of magnitude smaller than those in $d=2$ for all particle sizes. 
These higher average barriers are a first indication of the origin of the different dynamics in the two- and three-dimensional supercooled liquid. Nonetheless, the majority of barriers are too large to be relevant for structural relaxation. Even for the smallest particles in $d=3$, we find on average that $\overline{E}_\mathrm{b}\approx 10^2k_\mathrm{B}T$ around the mode-coupling point, corresponding to a Kramers escape time that far exceeds the structural relaxation time ($\tau_\mathrm{K} \approx 10^{40}$, $\tau_\alpha \approx 10^4$). 

\begin{figure}
    \centering
    \includegraphics[width = \linewidth]{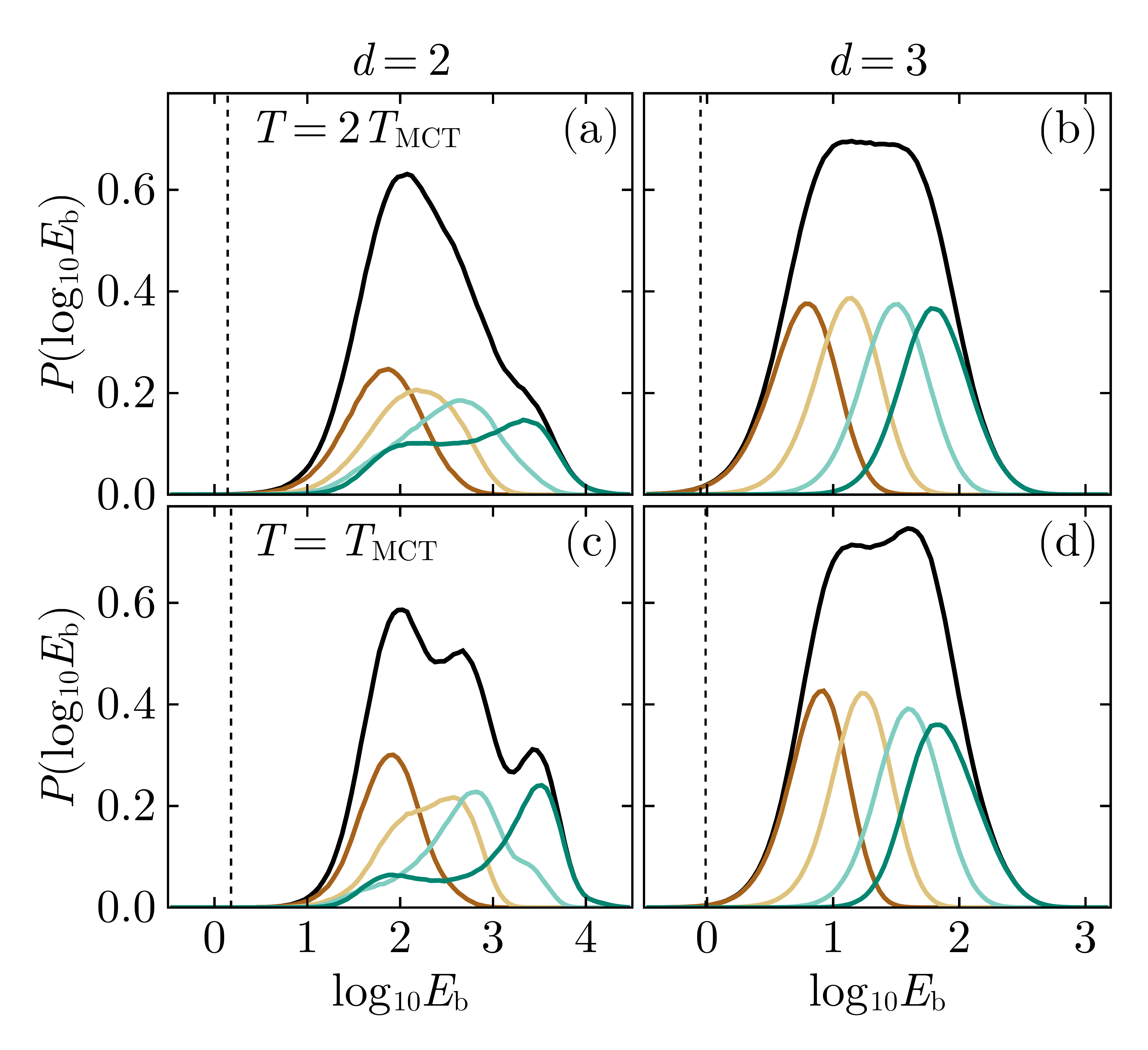}
    \caption{The distribution of the logarithm of the barrier heights in two and three dimensions resolved for different particle sizes for $T = 2T_\mathrm{MCT}$ (a, b) and $T = T_\mathrm{MCT}$ (c, d). The vertical dashed lines indicate the barrier heights $E_\mathrm{b}^\alpha = k_\mathrm{B}T\ln{\tau_{\mathrm{B}}}$ corresponding to a Kramers escape time equal to the structural relaxation time, as measured by the bond-breaking correlation function $C_\mathrm{B}(t)$. Legend colors are the same as in Fig.~\ref{fig:2} and the black line indicates the distribution irrespective of particle size.}
    \label{fig:5}
\end{figure}

To uncover the cases that are relevant to structural relaxation, we investigate the full distribution $P(\log_{10}(E_\mathrm{b}))$ of barrier heights. This distribution is shown in Fig.~\ref{fig:5}, where we present both the diameter-resolved (colored) and total (black) distributions of $\log_{10}(E_\mathrm{b})$ for $T=2T_\mathrm{MCT}$ and $T=T_\mathrm{MCT}$ for the two- and three-dimensional system. The distributions reveal that small particles consistently exhibit lower barriers compared to larger particles across all temperatures studied. This holds both in two and three dimensions and is consistent with the average energy barrier heights shown in Fig.~\ref{fig:4}. However, in $d=2$, at low temperatures, the total distribution of barrier heights shows a distinct separation, as indicated by the presence of three peaks [Fig.~\ref{fig:5}(a)]. This suggests that the particles are divided into three groups associated with different cage structures. Although the groups are generally correlated with particle size, the relationship is not perfect as some particles of the largest size quartile (green curve) are observed in the group with the lowest barrier heights. We believe this trichotomy is a reflection of the ``compositional ordering" observed and characterized by \citet{tong2023emerging}, who also identified three emergent particle subpopulations with distinct structural properties. These effects weaken as the temperature is increased, as indicated by the loss of multiple peaks for higher $k_\mathrm{B}T$ in Fig.~\ref{fig:5}(c). Interestingly, in $d=3$ [Figs.~\ref{fig:5}(b, d)], this separation is not observed since the distributions remain unimodal, even for the lowest temperatures in our data set. While the precise microscopic origins of this separation in $d=2$ are unclear, we believe it is due to a complex interplay between the non-additive nature of the system and its dimensionality. We have checked that the observed trichotomy persists in two-dimensional systems at higher polydispersity degrees and even transforms into a quadrichotomy at the highest polydispersities investigated (see SI).

\begin{figure}
    \centering
    \includegraphics[width = \linewidth]{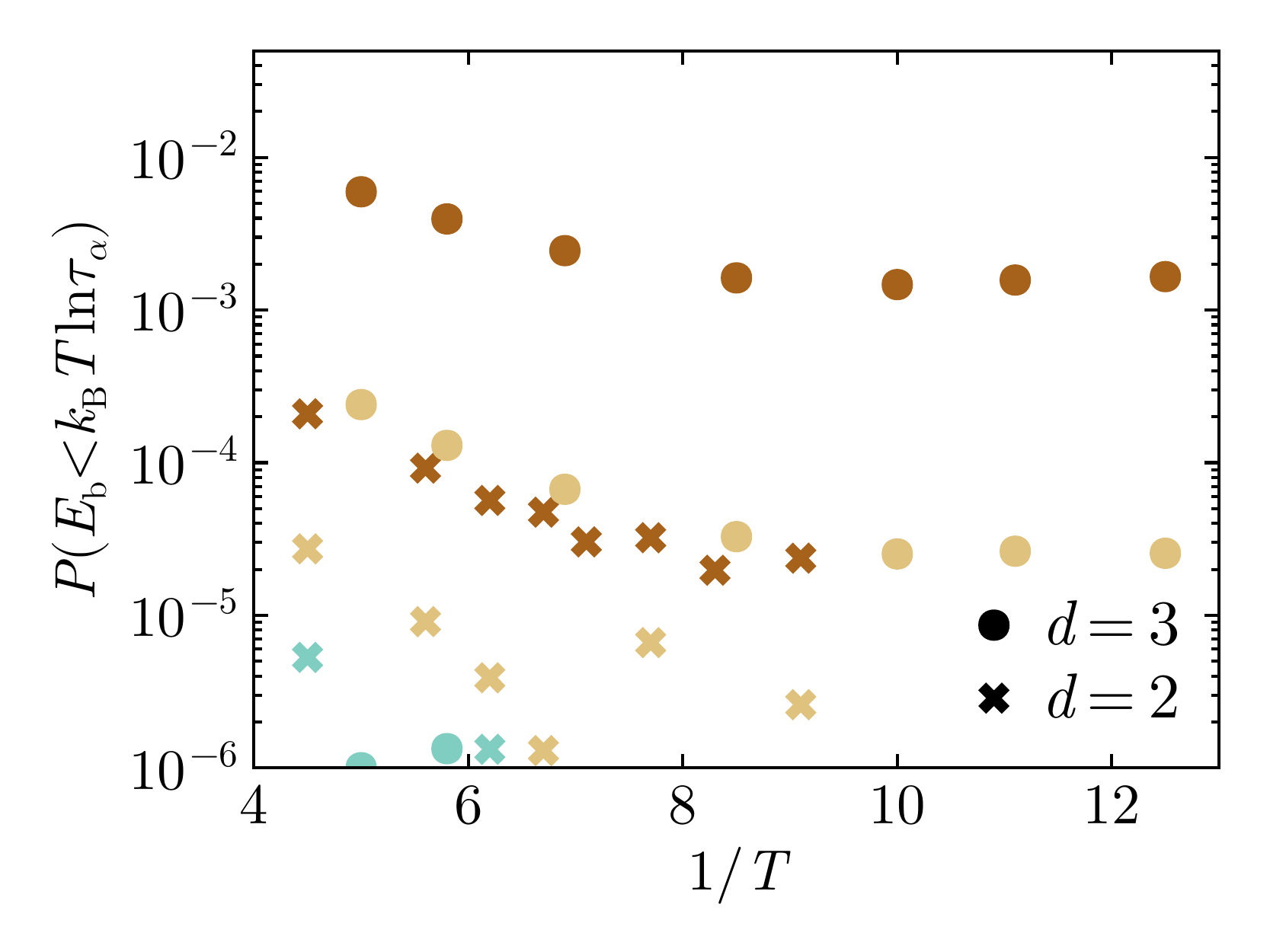}
    \caption{The fraction of tagged particles for which the estimated escape times are shorter than the bond-breaking relaxation time as a function of the inverse temperature. We show the data separately for each of the four particle-size bins and in two- and three-dimensional systems. Missing data points indicate that no particles satisfying the criterion have been found.}
    \label{fig:6}
\end{figure}

Given that the vast majority of the local energy barriers are too high to be relevant for structural relaxation, we seek to identify situations that could be relevant for relaxation. To this end, we search for cages for which the Kramers escape time is smaller than the structural relaxation time of the supercooled liquid ($\exp(-E_\mathrm{b}/k_\mathrm{B}T) < \tau_\alpha$). In Fig.~\ref{fig:6}, we show the fraction of cages that satisfy this condition. Clearly, small particles (brown symbols) obey this criterion much more frequently than larger ones (yellow and green symbols), both in two and three dimensions. Thus, in general, hopping events are more accessible to small particles than to large ones. This finding is also fully consistent with the behavior of single-particle trajectories in three dimensions reported in  Fig.~4 of Ref.~\cite{pihlajamaa2023influence}. 
However, our data in Fig.~\ref{fig:6} also reveal a clear difference across dimensions. Specifically, the probabilities/fractions $P(E_\mathrm{b}< k_\mathrm{B}T\ln \tau_\alpha)$ in two and three dimensions differ by almost two orders of magnitude. This suggests that the type of relaxation events captured by the Kramers model (colloquially: `single-particle hopping') is considerably more frequent in three dimensions than in two. While this minimal approach to localized cage escapes does not necessarily correspond to a genuine physical process in supercooled liquids, our results nonetheless inform why small particles progressively dominate the structural relaxation in $d=3$ but not in $d=2$. 

Intuitively, we can understand these dimensional differences by considering an analogy with a jammed hard-sphere system. In three dimensions, such a system always has voids that percolate, whereas its two-dimensional analogue does not show void percolation. Therefore, a sufficiently small particle placed in the voids of the respective systems must be localized in $d=2$ but not in $d=3$. While this is only a limiting picture, it nonetheless illustrates the dimensional differences found in our single-particle model, as well as the presence ($d=3$) and absence ($d=2$) of dynamical decoupling in the full structural relaxation of the supercooled liquid.

\section{Conclusion}

We have resolved a recent conundrum on the dynamics of a polydisperse model glassformer, which is used ubiquitously throughout the field of theoretical and computational glass physics. Specifically, whereas in three dimensions the relaxation dynamics of this system is strongly particle-size dependent \cite{pihlajamaa2023influence}, the two-dimensional dynamics were reported \cite{tong2023emerging} to become size-independent upon supercooling. Our work reveals that the latter observation is caused by artifacts induced by a standard Merwin-Wagner fluctuation correction procedure.
The increasing dynamical similarity  between small and large particles in 2d disappears if different -- and \textit{a priori} equally suitable -- dynamic correlation functions are used. Specifically, depending on the choice of correlation function, we find either a constant ratio or even a modest growth in the relative mobilities of differently sized particles. In all cases, however, this dynamical particle-size dependence remains significantly smaller in two dimensions than in three.

To gain a more complete picture of the polydispersity-induced differences between dimensions, we have analyzed particle-size-dependent aspects of dynamic heterogeneity in both 2d and 3d. By conducting a cluster analysis of the mobile particles in time, we find that structural relaxation is initiated by string-like excitations of mostly small particles. This holds in both 2d and 3d. However, the further growth of these mobile clusters shows a marked dependence on dimensionality. In three dimensions, mobile regions grow by facilitating neighboring small particles, culminating in sparse mobile clusters with an overrepresentation of small particles. By contrast, in two dimensions, the growth of mobile domains yields compact clusters that are indiscriminate with respect to particle size.

We explain these observed differences between 2d and 3d by analyzing the local energy landscapes experienced by  particles of different sizes. 
In particular, we have calculated the full distributions of minimal energy barriers as a function of particle size and temperature. A key finding is that the energy barriers in 2d are significantly larger than those in 3d. Within the framework of the reduced Kramers model, this suggests that dynamics in the 2d system must necessarily involve cooperative motion, as the large energy barriers prevent isolated single-particle events. In contrast, the smaller energy barriers in the 3d system allow for single-particle activated events to occur more readily, which can be further rationalized by a void-percolation argument.
Moreover, we find a pronounced size dependence of the energy barriers: the barriers for smaller particles are much lower than those for larger particles, both in 2d and 3d. This explains the preferential involvement of smaller particles in the initial stages of relaxation in both dimensions. 
Together, these findings provide a clear explanation of the dimensional and size-dependent differences in particle dynamics, shedding light on the fundamental processes governing relaxation in highly-polydisperse supercooled liquids.

\section{Methods}

\subsection{Simulations}

We study a recently developed polydisperse mixture introduced by Berthier and coworkers \cite{ninarello2017models, scalliet2022thirty}. It consists of $N$ particles that interact through the pair potential
\begin{equation}
    u(r_{ij}) = \epsilon \left(\frac{r_{ij}}{D_{ij}}\right)^{-12}+c_4\left(\frac{r_{ij}}{D_{ij}}\right)^4+c_2\left(\frac{r_{ij}}{D_{ij}}\right)^{2}+c_0,
\end{equation}
for $r_{ij}/D_{ij}<r_c$ and $u(r_{ij})=0$ otherwise.
The constants $c_0=-28\epsilon/r_c^{12}$, $c_2=48\epsilon/r_c^{14}$, and $c_4=-21\epsilon/r_c^{16}$ ensure that the potential is continuous and twice differentiable at the cutoff radius $r_c$. The pair diameter is determined by $D_{ij}= (D_i+D_j)(1 - \zeta|D_i-D_j|)/2$, where the nonadditivity parameter $\zeta=0.2$ inhibits crystallization through demixing. The pair diameters are drawn from a polydispersity distribution $P(D)=A/D^3$, for $D\in[D_{\mathrm{min}},D_\mathrm{max}]$, where $A=D_\mathrm{min}D_\mathrm{max}/(D_\mathrm{max}-D_\mathrm{min})$, and $D_\mathrm{min} = D_\mathrm{max}/( 2 D_\mathrm{max}-1)$. The latter choices ensure that the mean diameter is unity and the distribution is normalized. We set the maximal size ratio equal to $D_\mathrm{large}/D_\mathrm{small}=2.219$. This distribution can be divided into four quartiles, each with an equal number of particles. The corresponding bin edges are $D =$ [0.725, 0.811, 0.935, 1.144, 1.610]. We average over particles within each bin to compute particle-size resolved quantities.

To quantify the degree of polydispersity, one can introduce a set of polydispersity indices 
    \begin{equation}
        \delta_x = \sqrt{\frac{\langle x^2 \rangle - \langle x \rangle^2}{\langle x \rangle^2}}
    \end{equation}
where $x$ is an observable which varies for particles of different sizes. The conventional polydispersity index uses $x=D$, the diameter of the particles. This leads to a dimensionally independent polydispersity index of $\delta_D = 23\%$ for the mixtures considered in this work. We must note however that the volumetric polydispersity, where $x$ is the volume (or the area in 2d) of the particles is dimensionally dependent. In particular, we find that the volumetric polydispersity of the two- and three-dimensional mixtures are 48\% and 75\% respectively. Thus, the present three-dimensional system is significantly more polydisperse by particle volume than its two-dimensional counterpart. We have verified that the conclusions presented in this work are independent of this difference by simulating a two-dimensional system with a volumetric polydispersity of 91\%, well exceeding that of the three-dimensional system considered here.

We equilibrate the system of $N=1200$ particles in dimensions $d=2$ and $d=3$ at number density $\rho=1$ at constant temperature $T$ using the swap Monte Carlo algorithm for at least 100 relaxation times. From every equilibrated configuration, we start a molecular dynamics simulation in the microcanonical ensemble (NVE) for production. We obtain correlation functions by averaging over all independent runs.

\subsection{Correlation functions}

For identifying neighbors, we mostly follow \citet{scalliet2022thirty}. Specifically, for the bond-breaking function, we use the criterion $r_{ij}/D_{ij}<r_c$, with the cut-off $r_c = 1.7$ for $t>0$, and $r_c=1.3$ and $r_c=1.485$ in $d=2$ and $d=3$ at $t=0$ only. Decreasing the cut-off at $t=0$ removes the effect of cage-rattling and thereby the plateau from the bond-breaking function because bonds are considered broken only when particles have moved distances exceeding the first minimum of the pair-correlation function. We use the same neighbor criteria for the cage-corrected intermediate scattering function. For the bond-orientational functions, we use the $t=0$ values of $r_c$ at all $t$ (deviating from \citet{scalliet2022thirty}). We have verified that different criteria for identifying neighbors do not qualitatively affect the results for the correlation functions (other than perhaps reintroducing the plateau for $C_\mathrm{B}(t)$). Additionally, we have verified that for different choices of spherical harmonic orders $m$ (such as for $Y_{64}$, or $Y_{61}$) in the definition of $\Psi_j(t)$ in $d=3$, the results remain qualitatively unchanged. For the overlap function, we set $a=0.5$, which is bigger than the typical cage size to probe cage-escape dynamics, and for the intermediate scattering functions, we investigate the wave numbers $|\textbf{k}|=6.7$ in $d=2$ and $|\textbf{k}|=7.0$ in $d=3$, corresponding to the peaks of the respective static structure factors.

For the bond orientational correlation function, we use the function $\Psi_i(t)$ to characterize the local six-fold orientational order of particle $i$ at time $t$. In this work, we use 
\begin{equation}
        \Psi_j(t) = \frac{1}{n_j(t)}\sum_{l=1}^{n_j(t)}\begin{cases}
            e^{6i \phi_{lj}(t)}\qquad\qquad\,\, \mathrm{ if }\ d = 2, \\ 
            Y_{66}(\theta_{lj}, \phi_{lj}) \,\ \qquad \mathrm{ if }\ d = 3,
        \end{cases}
    \end{equation}
where $n_j(t)$ is the number of neighbors of particle $j$ at time $t$, $\theta_{lj}$ and $\phi_{lj}$ are the polar and azimuthal angles, and $Y_{lm}$ are the spherical harmonics.

\subsection{Mobile clusters}

We identify mobile particles by the criterion $C_\mathrm{B}^i(t) < 0.5$, where the bond-breaking order parameter $C_\mathrm{B}^i(t) = n_i(t; 0)/n_i(0)$ quantifies the fraction of the neighbors of particle $i$ that remain a neighbor at time $t$. Here, whether particles are neighbors at any time is determined by a Voronoi tessellation because distance criteria yielded spatial artifacts. In order to conduct the clustering procedure, we initially assign each mobile particle to its own cluster. Then, if two particles from two distinct clusters are neighbors by Voronoi tessellation, the corresponding clusters are merged. This is repeated until convergence. 

\subsection{Energy Barriers}\label{sec:methods}

In order to set up our reduced single particle systems, we identify neighbors of a chosen particle with a Voronoi tesselation in both two and three dimensions. To compute the energy barriers of this system, we use two methods to compute energy barriers that we call the variable angle method and the constant angle method. We describe these below.

Let $V(r, \theta)$ be the two-dimensional potential energy landscape in radial coordinates $r$ and $\theta$, defined with respect to the initial location of the central dynamic particle. Both the variable angle method and the constant angle method determine the potential barrier as the maximal potential energy encountered along some path $\theta(r)$ where $r$ increases monotonically. 

For the variable angle method, $\theta(r)$ is determined by the condition that for every given radial distance $r$ the angle $\theta$ minimizes $V(\theta, r)$. Because the angle $\theta(r)$ is not required to be continuous in $r$, this method might produce paths that the particle cannot physically traverse, and will therefore, in general, underestimate the true lowest energy barrier. 

The constant angle method finds the path defined by the constant angle $\theta$ where $\theta$ is chosen such that the encountered barrier is minimal. In general, because an optimal trajectory is not along a straight radial line, this method will overestimate the encountered barrier. In summary, we have
\begin{align}
   \theta(r) &= \mathrm{argmin}_\theta \,  V(r,\theta)  &\text{(variable angle)},\\
  \theta &= \mathrm{argmin}_\theta \left[\mathrm{max}_r  V(r,\theta)\right]  &\text{(constant angle)},
\end{align}
where $\mathrm{argmin}_x\,f(x,y)$ is equal to the value of $x$ at the global minimum of $f(x,y)$ for fixed $y$.

\begin{figure}[ht!]
    \centering
    \includegraphics[width=\linewidth]{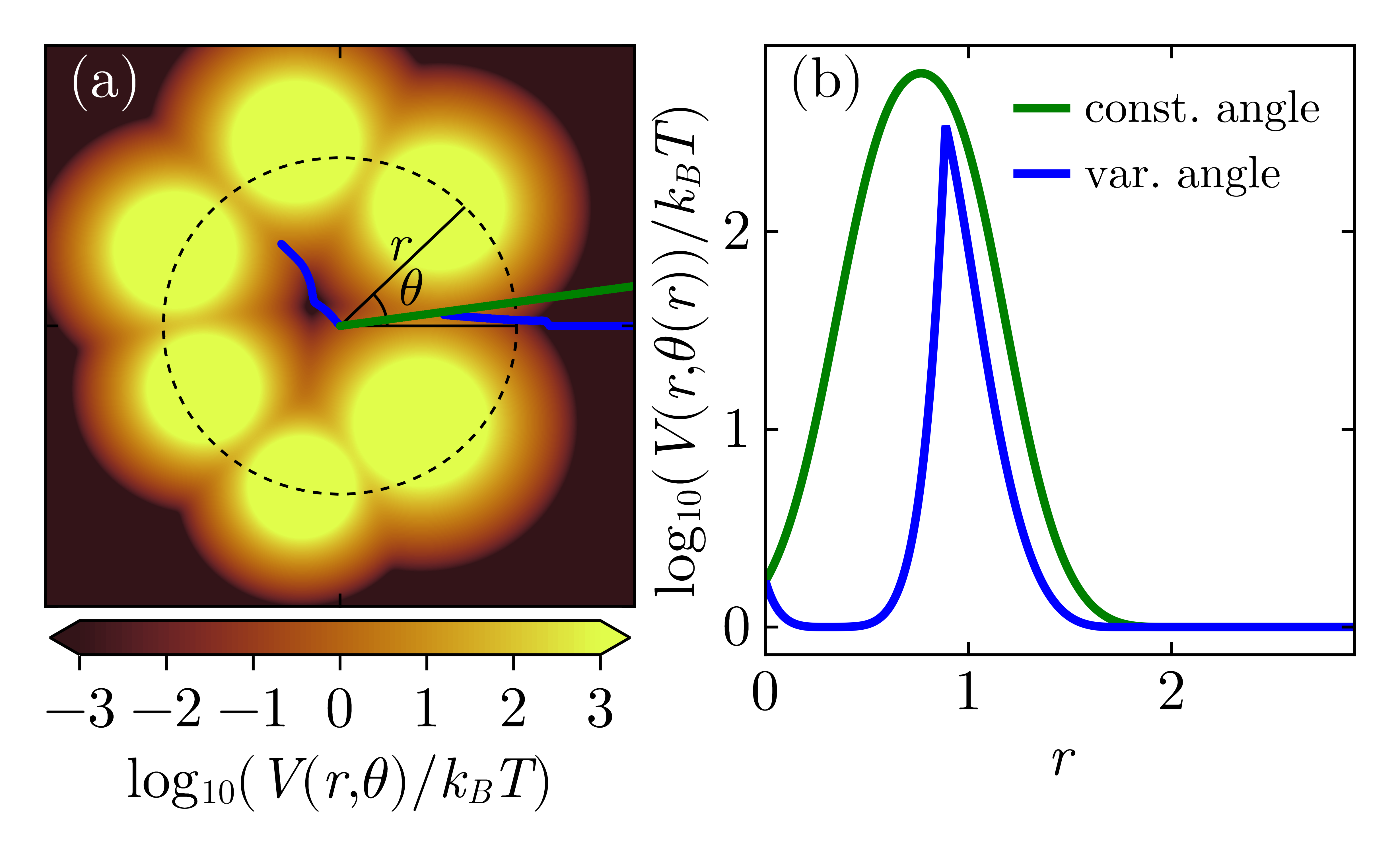}
    \caption{Two methods for determining the barrier height $E_\mathrm{b}$. Panel (a) shows the potential energy field of the mobile particle as a function of its position. The colors denote the logarithm of the potential energy. With blue and green lines, we show the two methods we use to determine the barrier height. The green line shows the constant angle method. The corresponding path is determined by the direction $\theta$ in which the encountered potential barrier is minimal. The blue line shows the variable angle method. The corresponding path $\theta(r)$ is determined by requiring that for every given radial distance $r$ the angle $\theta$ minimizes $V(r, \theta)$. Panel (b) shows the energy landscape as encountered along the two paths. The energy barriers encountered along these two paths provide upper and lower bounds for the minimal energy barrier the particle must overcome to escape.}
    \label{fig:7}
\end{figure}

An example of the paths obtained by both methods is shown in Fig.~\ref{fig:7}(a). For both methods, the potential is set to zero at the initial position of the caged particle, $V(r=0, \theta) = 0$. In Fig.~\ref{fig:7}(b), we show the potential energy encountered along these paths. Replacing $\theta$ by the solid angle $\Omega$, both methods are trivially generalized to three dimensions.

\begin{figure}[ht]
    \centering
    \includegraphics[width=0.8\linewidth]{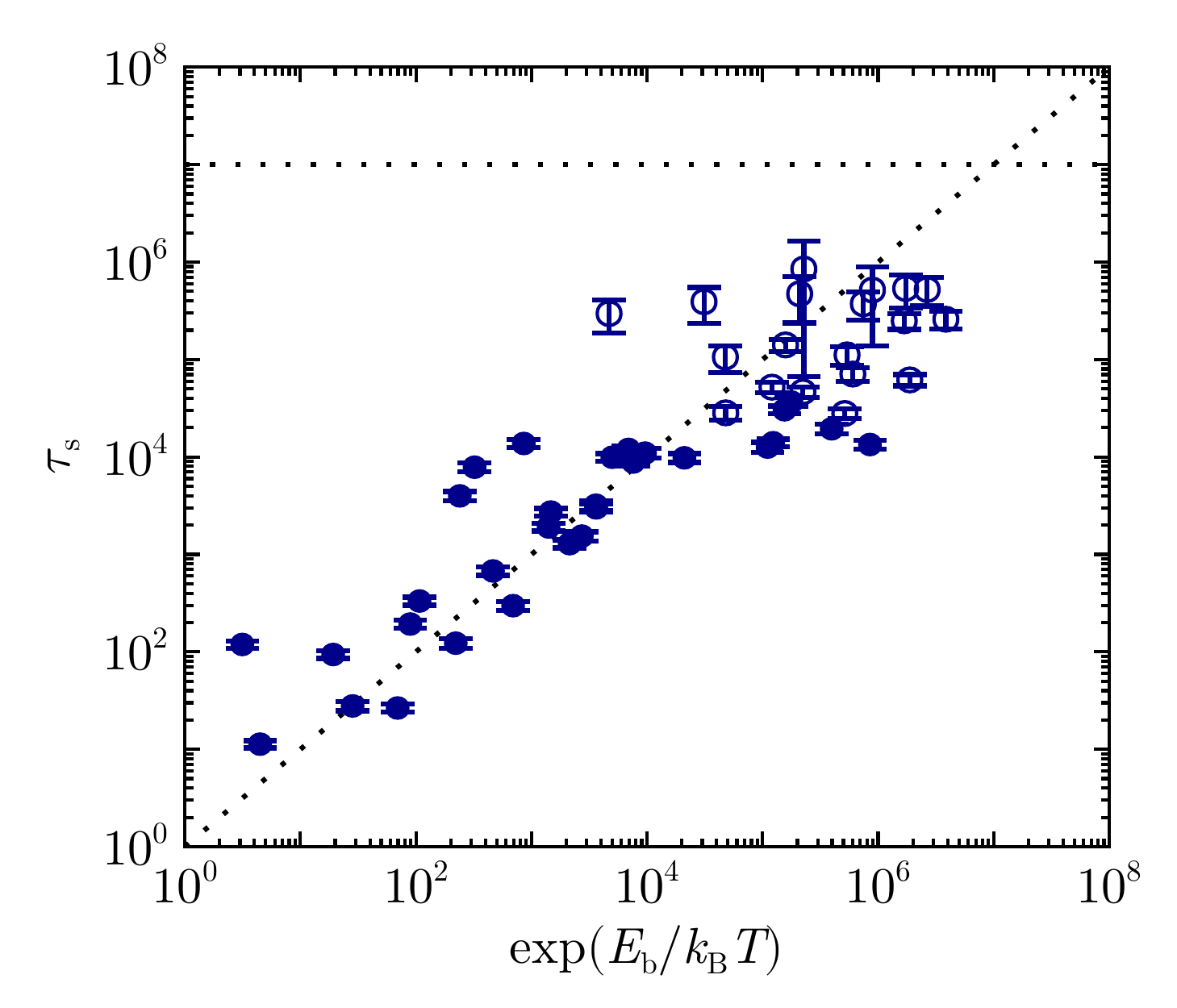}
    \caption{Average simulated escape time $\tau_\mathrm{s}$ compared to Kramers escape time $\tau_\mathrm{K}$ computed with the variable angle method. Each data point corresponds to an average over an ensemble of 100 simulations. The empty markers indicate cages for which not all particles escaped within our limit of $10^7$ time units (indicated by the horizontal dashed line). }
    \label{fig:3_2}
\end{figure}

To explicitly verify that the calculated barrier heights determine the escape time, we have simulated a sample of the particles in their respective pinned cages using Brownian dynamics with the adaptive Euler-Heun method \cite{kloeden1992stochastic}. We show in Fig.~\ref{fig:3_2} that our simulated single-particle systems indeed obey Kramers' law by comparing the simulated escape time $\tau_{\mathrm{s}}$ with the theoretical prediction $\tau_{\mathrm{K}}$. 

\section*{Acknowledgements}

IP, CCLL and LMCJ acknowledge funding from a Vidi grant from the Dutch Research Council.


\bibliography{apssamp}

\end{document}